\documentclass{article}

\usepackage{arxiv}
\usepackage{amsmath,amsfonts,amssymb,amsthm}
\usepackage{mathtools}
\usepackage[utf8]{inputenc} 
\usepackage[T1]{fontenc}    
\usepackage{hyperref}       
\usepackage{url}            
\usepackage{booktabs}       
\usepackage{amsfonts}       
\usepackage{nicefrac}       
\usepackage{microtype}      
\usepackage{lipsum}
\usepackage{graphicx}
\usepackage{textcomp}
\usepackage{stfloats}
\usepackage{url}
\usepackage{verbatim}
\usepackage{graphicx}
\usepackage{cite}
\usepackage{algorithmic}
\usepackage{algorithm}
\usepackage{array}
\usepackage{subfigure}
\usepackage{textcomp}
\usepackage{stfloats}
\usepackage{url}
\usepackage{verbatim}
\usepackage{graphicx}

\graphicspath{{./figure/}}
\DeclareGraphicsExtensions{.pdf,.jpeg,.png,.jpg}

\title{Achieving Stability and Optimality: Control Strategy for a Wind Turbine Supplying an Electrolyzer in the Islanded Storage-less Microgrid}

\author{
  Bosen Yang \\
  The Department of \\
  Electrical Engineering\\
  Tsinghua University\\
  Beijing, China \\
  \texttt{ybs22@mails.tsinghua.edu.cn} \\
  \And
 Kang Ma \\
 China Datang Technology \\
 Innovation Co., Ltd\\
 Xiong'an, China \\
  \texttt{makang1984@gmail.com} \\
 \And
  Jin Lin \\
  The Department of \\
  Electrical Engineering\\
  Tsinghua University\\
  Beijing, China \\
  \texttt{linjin@mails.tsinghua.edu.cn} \\
 \And
 Mingjun Zhang \\
 State Grid Ningxia \\
 Electric Power Co.,Ltd\\
  Ningxia, China \\
  \texttt{zmj17zhangmj@163.com} \\
 \And
  Qiwei Duan \\
  China Datang Technology \\
  Innovation Co., Ltd\\
  Xiong'an, China \\
  \texttt{Dqwforfly@163.com} \\
  \And
  Zhendong Ji \\
  Anhui Usem \\
  Technology Co., Ltd\\
  Anhui, China \\
  \texttt{zhendong\_ji@njust.edu.cn} \\ 
  \And
  Zhi Liu \\
  Sinovel Wind Group Co., Ltd.\\
  Beijing, China \\
  \texttt{liuzhi@sinovel.com} \\
  \And
 Yonghua Song \\
 The Department of Electrical\\
 and Computer Engineering\\
 University of Macau\\
 Macau China\\ 
 \texttt{yhsong@tsinghua.edu.cn} \\
}

\begin{document}
\maketitle

\begin{abstract}
  Wind power generation supplying electrolyzers in islanded microgrids is an essential technical pathway for green hydrogen production, attracting growing attention in the transition towards net zero carbon emissions. Both academia and industry widely recognize that islanded AC microgrids normally rely on battery energy storage systems (BESSs) for grid-forming functions. However, the high cost of BESS significantly increases the levelized cost of hydrogen (LCOH), compromising economic feasibility. To address this challenge and reduce the LCOH, this paper focuses on a wind turbine (WT) supplying an electrolyzer in a storage-less microgrid and identifies a unique characteristic that challenges the conventional understanding of this microgrid: active power is coupled with microgrid voltage rather than frequency, the latter being entirely decoupled from active power balance. Based on this unique characteristic, this paper develops a new control strategy that maintains power balance, stabilizes the voltage and frequency, and maximizes hydrogen production. The effectiveness of the control strategy is validated through case studies conducted in Matlab/Simulink, especially its capability to maintain stability while maximizing hydrogen production under various conditions. 
  \end{abstract}


\section{Introduction}
\subsection{Background and Motivation}
The global wind industry has been experiencing rapid growth. In 2023, a record 117 GW of wind capacity was added, bringing the total global wind capacity to 1021 GW \cite{ref1}. According to the International Energy Agency's Net Zero Emissions by 2050 Scenario (NZE Scenario), achieving net-zero carbon emissions requires annual wind installations to reach 320 GW by 2030, with cumulative global wind capacity expected to rise to 2.75 TW by the same year \cite{ref2}. However, despite these ambitious targets, one of the primary challenges hindering wind power expansion is the increasingly limited hosting capacity of electrical grids \cite{ref3}. Take China as an example: in 2023, the average wind curtailment rate (AWCR) in some congested regions exceeded 6\% \cite{ref4}. In addition to curtailment challenges, high wind power penetration poses further difficulties, especially in grid stability \cite{ref5}. Therefore, careful planning is essential before integrating numerous wind turbines with the grid \cite{ref6}.\par
Wind-powered hydrogen production offers a promising solution for accommodating renewable energy and producing the much-needed green hydrogen \cite{ref7}. As a critical component of the transition to net-zero carbon emissions, green hydrogen has diverse applications across industries, including refining, chemical production (e.g., ammonia and methanol synthesis), and steel manufacturing \cite{ref8}. Green hydrogen remains scarce despite its potential: global hydrogen demand reached 97 Mt in 2023, yet only 1\% was produced from renewable sources \cite{ref8}. A significant production scale-up is essential to meet the 65 Mtpa of green hydrogen required by 2030 under the NZE Scenario \cite{ref2}. This paper investigates an islanded microgrid with wind power generation supplying an electrolyzer.
\subsection{Literature Review}
Systems of WTs supplying electrolyzers can be classified into two types according to their relations with the grid: grid-connected \cite{ref9,ref10,ref11,ref12,ref13,ref14} and islanded \cite{ref15,ref16,ref17,ref18,ref19,ref20,ref21}. Kinds of literature on these two types are reviewed:\par
For grid-connected wind-to-hydrogen systems, substantial research has focused on leveraging the flexibility of electrolyzers to mitigate wind power curtailment and provide grid support. References \cite{ref9,ref10,ref11} explore using hydrogen as an energy storage medium to smooth the power delivery to the grid, significantly reducing AWCR. Reference \cite{ref12} investigates dynamic frequency regulation enabled by a grid-scale alkaline electrolyzer (AEL) plant. Reference \cite{ref13} indicates that the electrolyzers' flexible and fast ramping capability allows a wind electrolysis joint system to provide secondary frequency regulation. Reference \cite{ref14} introduces a nonlinear model predictive control approach for the AEL to offer ancillary services. \par
However, grid-connected systems have drawbacks: Firstly, to achieve the ambitious green hydrogen production goals outlined in the IEA's NZE scenario, relying solely on the byproduct hydrogen from wind-hydrogen systems that primarily provide ancillary services to the grid is inadequate. Secondly, large-scale grid-connected green hydrogen productions inevitably congest the grid, and peak-shaving would become inevitable \cite{ref22}. Thirdly, using grid power for hydrogen production, even though at a small percentage, means that the hydrogen is not 100\% green because the grid power is not 100\% from renewable sources \cite{ref22}. Specifically, the grid power is likely to be “grey” when the renewable power is low and loads are at peak. To overcome the shortcomings, islanded wind-to-hydrogen microgrids are a promising solution and a future trend.\par
Islanded microgrids can be further divided into DC microgrids \cite{ref15,ref16,ref17,ref18} and AC microgrids \cite{ref19,ref20,ref21}. As detailed in \cite{ref15,ref16,ref17,ref18}, islanded wind-powered hydrogen production microgrids based on DC microgrids typically have permanent magnet synchronous generators connected to electrolyzers via AC/DC converters cascaded by DC/DC converters. However, this topology is constrained by the limited transmission capacity and the high cost of DC power transmission, restricting its application to short-distance scenarios. Conversely, islanded AC microgrids presented in \cite{ref19,ref20,ref21} are technically feasible for wind-powered hydrogen production systems. Academia and industry broadly recognize that implementing islanded AC microgrids typically depends on BESS for grid-forming \cite{ref21}. Unfortunately, incorporating BESS significantly increases the cost of green hydrogen production, undermining its market competitiveness. To address this challenge, this study proposes a novel control strategy for a WT supplying an electrolyzer in an islanded storage-less microgrid to address this issue. This study employs the Type-III wind farm based on a doubly-fed induction generator (DFIG).

 \subsection{Contributions}
 The main contributions of this paper are threefold:\par
 1) In pursuit of reducing the LCOH, this study identifies unique characteristics based on a WT supplying an electrolyzer in a storage-less islanded microgrid. The distinctive characteristics are the microgrid frequency and voltage stability mechanisms. The frequency is decoupled from active power in the specific topology of the storage-less islanded microgrid consisting of a WT supplying an electrolyzer.  This contrasts with the traditional principle of conventional power systems, where the frequency depends on the system's active power balance. Instead, maintaining a constant microgrid voltage achieves the active power balance.  \par
 2) This paper develops a new control strategy for an islanded AC microgrid consisting of a WT supplying an electrolyzer, maintaining 50Hz frequency and voltage stability without the BESS for grid-forming functions. This is achieved by harnessing the WT and electrolyzer's flexibility and rapid response capabilities across a broad power range. The control strategy assigns the frequency regulation task to the DFIG-side converter and the voltage regulation task to the electrolyzer-side converter. This work eliminates the need for BESS, thus saving the cost of green hydrogen production significantly while maintaining a stable and secure microgrid operation. \par
 3) An uncontrolled rectifier (UR) supplied by a current source (CS) is modeled. In the microgrid topology investigated in this study, the DFIG operates as a CS, while the steady-state characteristics of the electrolyzer are modeled as resistive loads. Notably, the system lacks voltage sources (VSs). Thus, the mathematical relationship between the voltage and current given by this paper deviates significantly from the conventional models commonly found in power electronics literature.\par
 The remaining structure of this paper is organized as follows: Section II introduces the topology of the islanded microgrid, which consists of a WT supplying an electrolyzer with all necessary converters but without grid-forming BESS. Section III elucidates the critical basis for the existence of the stability operation point and outlines the optimal scheduling model. Section IV details the proposed control strategy for the islanded microgrid, considering two distinct operational modes. Section V validates the proposed approach through case studies conducted in Matlab/Simulink. Finally, the conclusions are summarized in Section VI.

 \section{The Topology for an Islanded Microgrid: a Wind Turbine Supplying an Electrolyzer without Storage}
 \begin{figure}[t]
 \centering
 \includegraphics[width=6in]{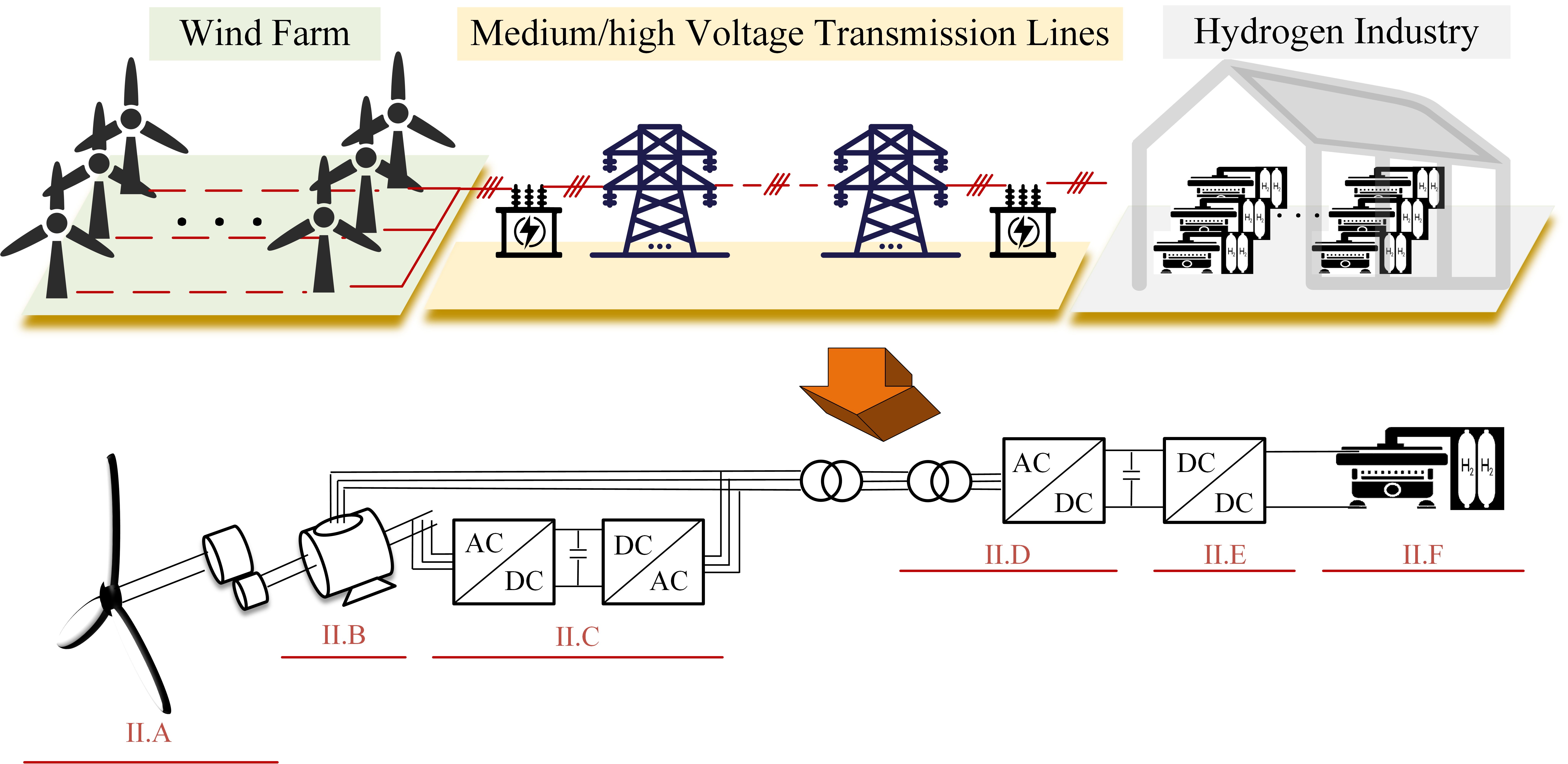}
 \caption{The topology of the islanded storage-less microgrid}
 \label{fig_1}
 \end{figure}
 Fig. \ref{fig_1} gives an overview of the islanded storage-less microgrid. A wind farm generates electricity, which is transmitted via medium/high voltage transmission lines to water electrolyzers for hydrogen production. Fig. \ref{fig_1} also presents a detailed electrical topology: a WT supplies an electrolyzer, accompanied by necessary converters without BESS. This paper focuses on an islanded microgrid as a single WT supplying a single electrolyzer (SWSE). This microgrid consists of the following key components: a WT, a DFIG, a back-to-back (B2B) converter, a 24-pulse uncontrolled rectifier (24-UR), and a 16-phase interleaved parallel buck converter (16-IPBC). Further details about this microgrid are discussed in subsequent sections.
 
 \subsection{WT Model}
 A WT model is adopted based on the aerodynamic principles governing the blade's behavior. The detailed model can be found in Reference \cite{ref21}.
 The maximum WT mechanical power is expressed as:
 \begin{equation}
   \label{eq1}
   P_m^{\max }=\frac{1}{2} C_p^{\max } \rho \pi R^2\left(\frac{\omega_m R}{\lambda_{\text {opt }}}\right)^3
 \end{equation}
 where $\rho$ is the air density; $R$ is the radius of the WT blade; $C_p^{\max }$ is the maximum power coefficient, which is a function of the tip speed ratio $\lambda$  and the blade pitch angle $\beta$ . $ \omega_m $  is the WT rotational angular frequency. 
 
 \subsection{DFIG Model}
 Several references introduce the general mathematical equations of asynchronous machines in the dq-frame \cite{ref23}. This paper employs sine-based Park transformation with the q-axis aligned with the a-axis at $\omega t=0$ . The voltage and flux linkage equations in the dq-frame are presented in Appendix \cite{ref24}.
 
 \subsection{B2B Converter Model}
 The B2B converter consists of two VS converters (VSCs) and a DC link between them. The VSC closer to the generator is the machine-side converter (MSC), and the other one is the line-side converter (LSC). The voltage equations of the converters in the dq-frame are detailed in \cite{ref25}.
 
 \subsection{Current Source Powered Uncontrolled Rectifier}
 \begin{figure}[b!]
   \centering
   \includegraphics[width=6in]{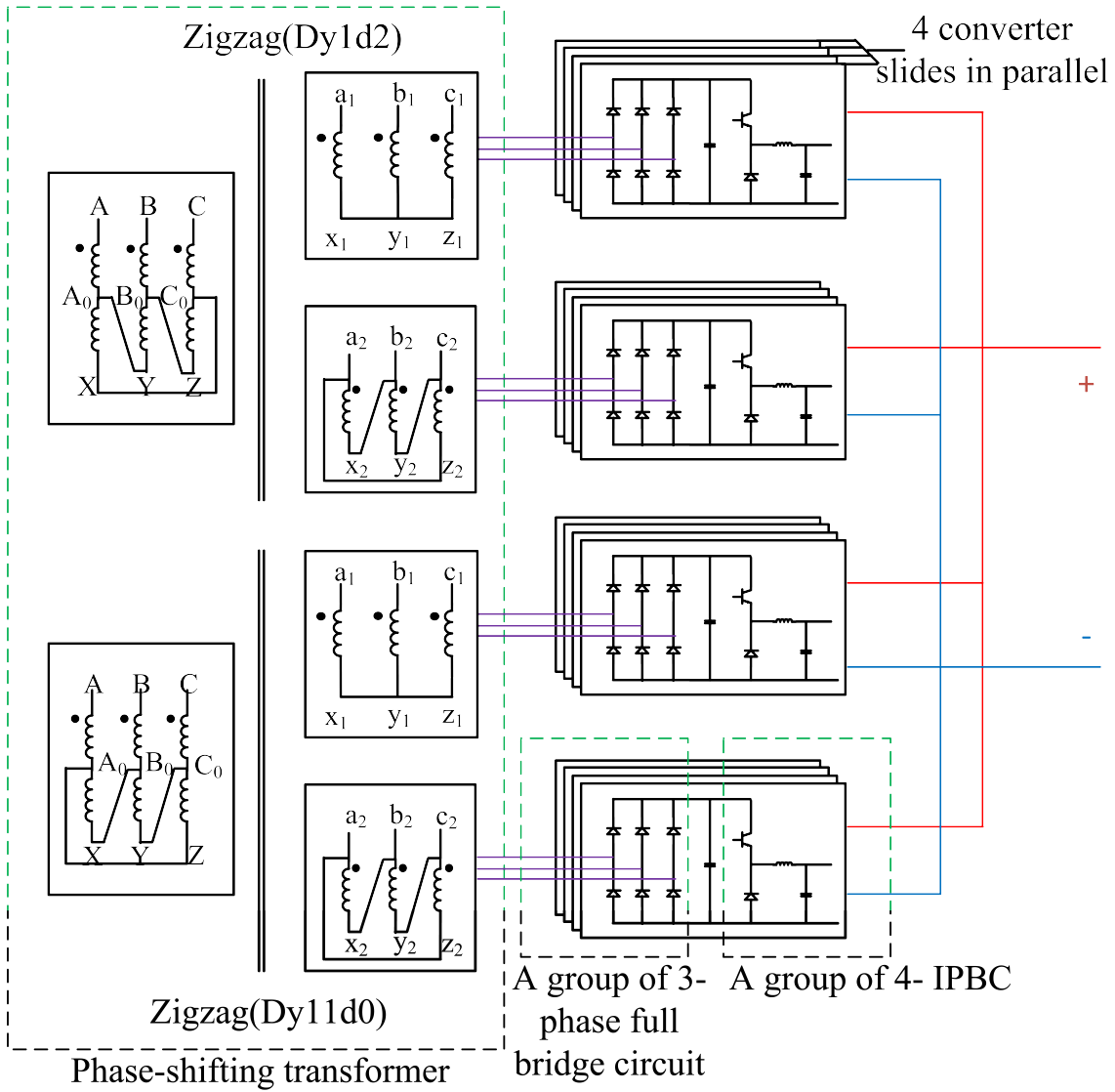}
   \caption{The topology of the 24-UR and the 16-IPBC}
   \label{fig_2}
 \end{figure}
   Fig. \ref{fig_2} illustrates the topology of the 24-UR and the 16-SIBC. The 24-UR comprises a phase-shifting transformer and four groups of 3-phase-full-bridge circuits \cite{ref26}. This study assumes the phase-shifting transformer winding connection follows (Dy1d2, Dy11d0). A converter slide shown in Fig. \ref{fig_2} consists of a 3-phase-full-bridge circuit and a buck converter; each group of 3-phase-full-bridge circuit has four circuits connected in parallel to increase the total current capacity.\par
\begin{figure}[t!]
     \centering
     \includegraphics[width=6in]{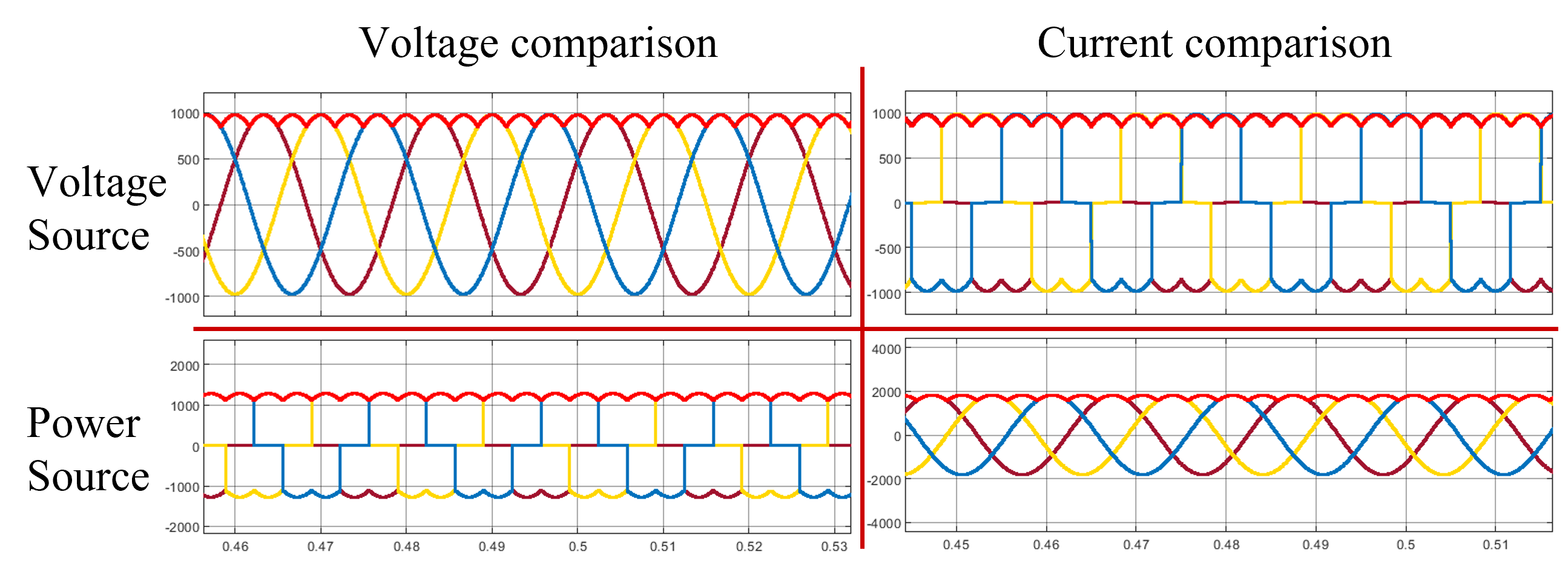}
     \caption{Comparison of the electrical characteristics of the UR when the AC side is a VS and a CS, respectively}
     \label{fig_3}
   \end{figure}

   Unlike existing references on uncontrolled rectifiers \cite{ref27}, this paper develops a configuration where the uncontrolled rectifier is powered by a CS rather than a VS. Fig. \ref{fig_3} compares the external characteristics of the rectifier port under two scenarios: the AC side being a VS versus a CS. In the former scenario, the AC voltage waveform is a sine wave controlled by the VS, and the line current waveform coincides with the DC load current when the corresponding valves are turned on; the DC voltage waveform is the envelope of the line voltage, and the DC current depends on the load.  In the latter scenario, the line voltage waveform mirrors the DC load voltage waveform when the corresponding valves are turned on, or zero if those valves are turned off. The DC current waveform is the envelope of the line current, and the DC voltage depends on the load. 
   Equations (\refeq{eq2}) - (\refeq{eq4}) compute the main electrical parameters.
   \begin{equation}
     \label{eq2}
     I_{d c}=\frac{3}{\pi} \int_{\frac{\pi}{3}}^{\frac{2 \pi}{3}} \sqrt{2} I_{a c} \sin (\omega t) d(\omega t)=1.35 I_{a c}
   \end{equation}
 
   \begin{equation}
     \label{eq3}
       U_{a c}=\sqrt{\frac{1}{2 \pi}\left[\frac{2 \pi}{3} U_{d c}^2+\frac{2 \pi}{3}\left(-U_{d c}\right)^2\right]}=0.816 U_{d c}
   \end{equation}
 
   \begin{equation}
     \label{eq4}
         U_{a c, 1}=\frac{\sqrt{6}}{\pi} U_{d c}=0.78 U_{d c}
   \end{equation}
  where $ I_{d c} $ and $ U_{d c} $ are the average values of output DC current and DC voltage, respectively; $ I_{a c} $ and $ U_{a c} $ are the RMS values of AC line current and AC line voltage, respectively; and $ U_{a c, 1} $ is the RMS value of the fundamental harmonic of the AC line voltage.
 
 \subsection{IPBC Model}
 In this study, a 16-IPBC is formed by connecting four groups of 4-IPBC in parallel, as shown in Fig. \ref{fig_3}. For simplicity, only the frequency domain model of the 4-phase interleaved parallel buck circuit derived by one of the authors Mingjun Zhang is used in the controller design, the detailed model is provided in Appendix \cite{ref24}. The relationship between input voltage $ u_{in} $ and output voltage $ u_{out} $ is given by:
 \begin{equation}
   \label{eq5}
   u_{out}=d u_{in}
 \end{equation}
 
 \subsection{AEL Model}
 Numerous studies discuss AEL models, including physics models \cite{ref28}, experience-based models \cite{ref29}, and data-driven models \cite{ref30}. This study focuses on the overall system control strategy and, thus, a precise but complex AEL model beyond the scope. Consequently, the AEL model is simplified to an internal resistance and a reverse VS.
 
 The AEL is modeled with a maximum ramping rate of 20\% load/s \cite{ref31}; however, from the actual test results of the AEL, a ramping rate of 5\% load/s is also challenging. Therefore, to make the simulation results closer to the exact engineering project, the ramping limitation is assumed to be 5\% load/s. The load range of AEL is 20\%-100\% \cite{ref32} for safety reasons. However, the load range can be extended to 10\%-100\% by some novel control method \cite{ref33}. In this study, the minimum load of an AEL is assumed to be 10\%.

 \section{Foundations of Stability and Steady-State Optimal Scheduling For the Microgrid}
 \subsection{The Frequency Stability Foundations}
 The concept of frequency in a 100\% converter-based microgrid differs fundamentally from that in a synchronous generator (SG)-based power system. In 100\% converter-based microgrids, frequency is a virtual value that can be flexibly regulated by the power electronics converter, independent of the power balance between sources and loads. Conversely, in SG-based systems, frequency corresponds to the synchronous speed of the generators and is inherently influenced by the active power dynamics of sources and loads. Recently, some researchers have taken notice of this issue and proposed frequency-fixed grid-forming inverter control methods \cite{ref34}.
 
\begin{figure}[t]
   \centering
   \includegraphics[width=6in]{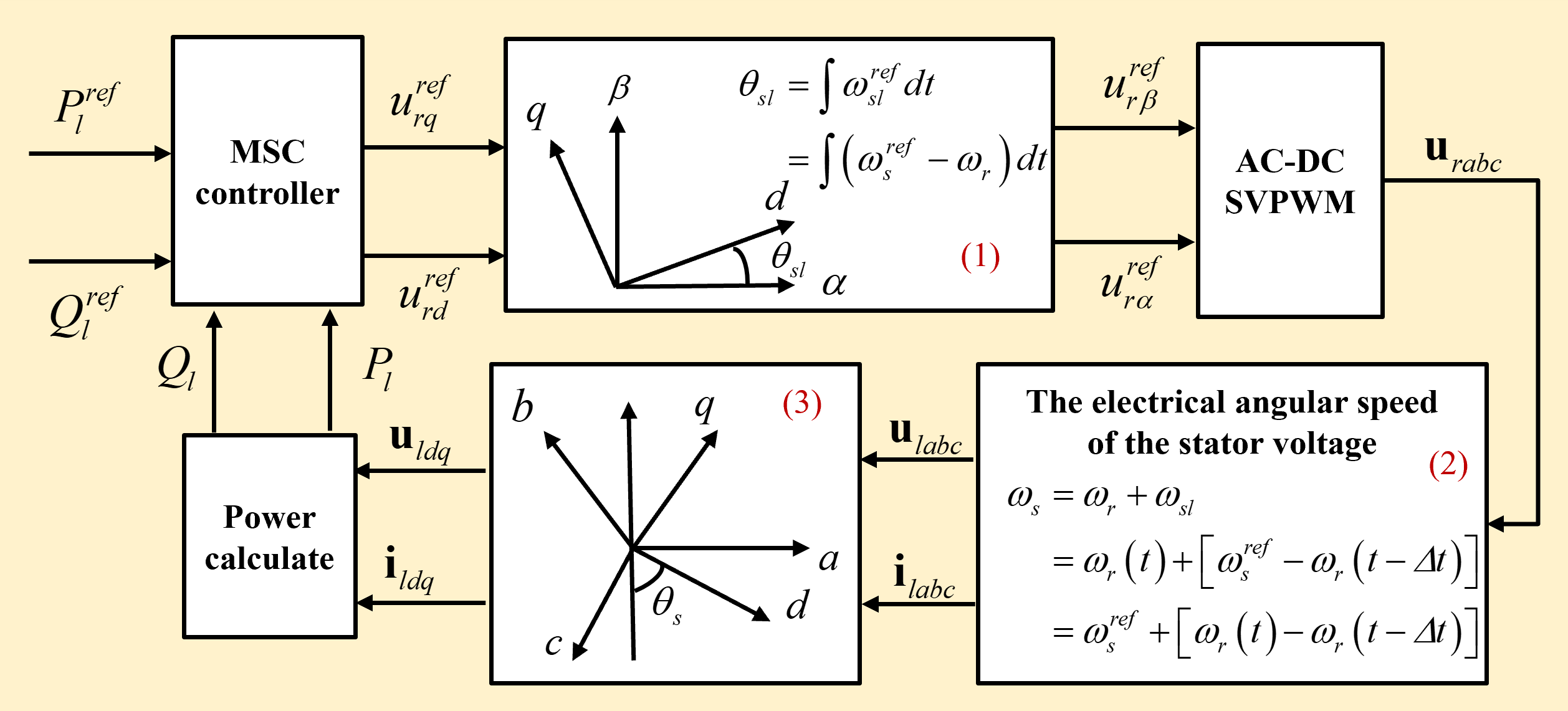}
   \caption{The basis of frequency stability}
   \label{fig_4}
 \end{figure}

 Fig. \ref{fig_4} illustrates the frequency control mechanism for the SWSE microgrid. The MSC controller takes the input variables: reference active/reactive power $P_l^{r e f}$ and $Q_l^{r e f}$, actual active/reactive power $P_l$ and $Q_l$ that are the outputs from the DFIG to the microgrid. The MSC controller outputs reference rotor voltages $u_{rd}^{r e f}$ and $u_{rq}^{r e f}$ in the $ {dq} $-frame in process (1). $\omega_s^{r e f}$ is the angular frequency at which the stator voltage is expected to operate. It can be set as any arbitrary value. In this case, it is set as $2 \pi \times 50$ rad/s. The slip angle $\theta_{s l}$ is obtained by integrating the difference between $\omega_s^{r e f}$ and $\omega_r$ over time. $\theta_{s l}$ serves as the reference angle for transforming the $ {dq} $-frame into the $ {\alpha\beta} $-frame. In this way, the reference rotor voltages $u_{r\alpha}^{r e f}$ and $u_{r\beta}^{r e f}$ in $ {\alpha\beta} $-frame embeds their angular frequency $\omega_{sl}$. The AC-DC converter generates the three-phase rotor voltage $u_{rabc} $ by applying space vector pulse width modulation. Block (2) indicates that the sum of the rotor angular frequency $\omega_r$ and the slip angular frequency $\omega_{sl}$ at $ t $  equals the stator angular frequency $\omega_s$ \cite{ref35}. However, under the premise that $\omega_{s l}(t)$ tracks $\omega_{s l}^{ref}(t)$ of the process (1), $\omega_{s l}(t)$ is derived from the rotor angular frequency $\omega_r(t-\Delta t)$ at the last sample time, so an error exists in $\omega_{s}$, as presented in block (2). In block (3), the line-side currents $\mathbf{i}_{labc}$ and voltages $\mathbf{u}_{labc}$ are sampled in the $ {abc} $-frame and then converted into $\mathbf{i}_{ldq}$ and $\mathbf{u}_{ldq}$ in the $ dq $-frame through Park transformation. The reference angular of Park transformation $\theta_{s}$ is obtained by the phase-locked loop (PLL). The error between the accurate frequency and the approximate frequency affects the amplitude of $\mathbf{u}_{ldq}$ and $\mathbf{i}_{ldq}$, which further affects the calculated power. $ P_l $ and $ Q_l $ are computed using the instantaneous power theory proposed by Akagi \cite{ref36}.  Finally, the MSC controller performs a closed-loop control, making $ P_l $ and $ Q_l $ track $ P_l^{ref} $ and $ Q_l^{ref} $, respectively.
 
 \subsection{The Voltage Stability Foundations}
 \begin{figure}[b]
   \centering
   \subfigure[]
   {
       \begin{minipage}[b]{.45\linewidth}
           \centering
           \includegraphics[width=3in]{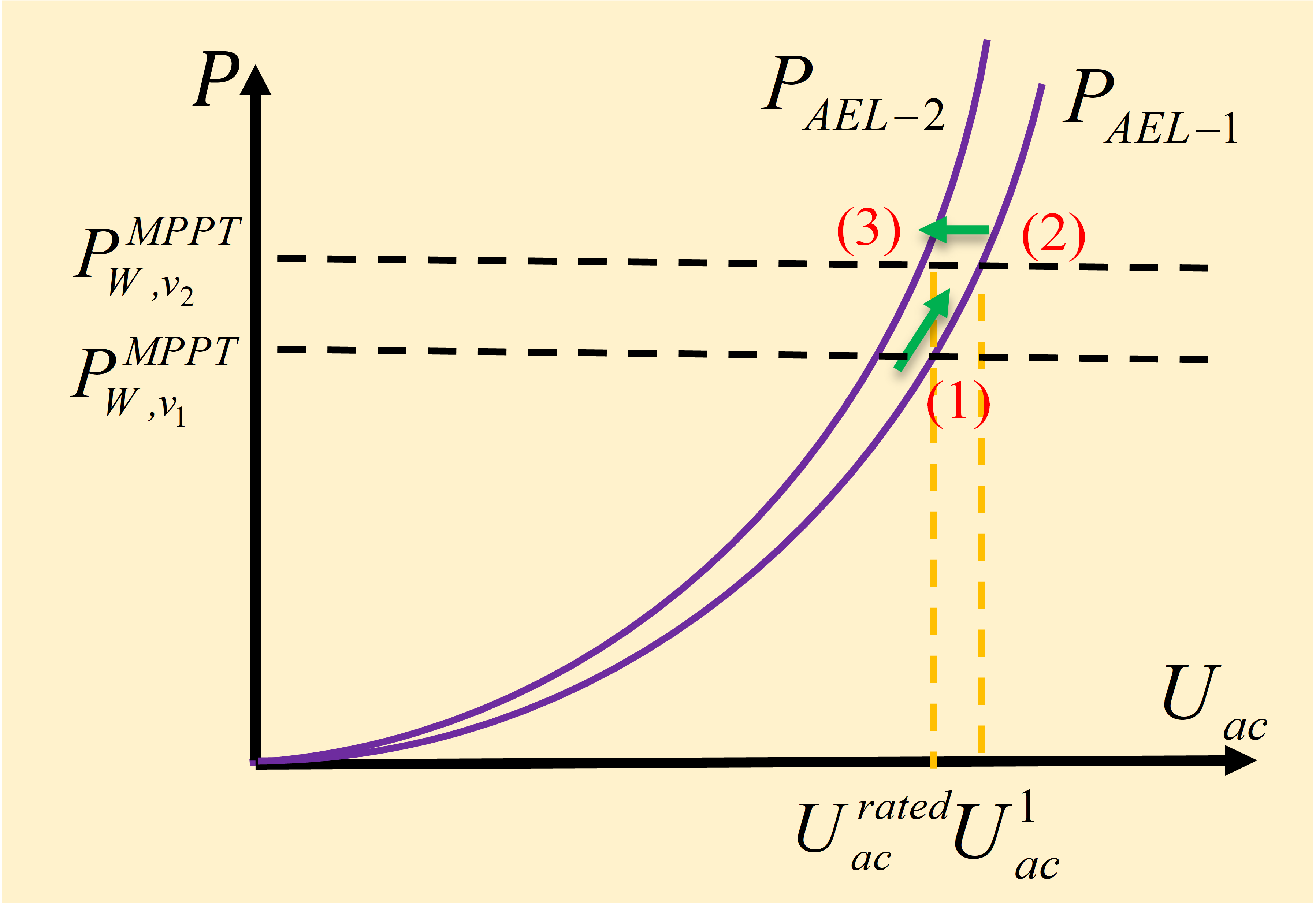}
       \end{minipage}
       \label{fig_5a}
   }
   \subfigure[]
   {
      \begin{minipage}[b]{.45\linewidth}
           \centering
           \includegraphics[width=3in]{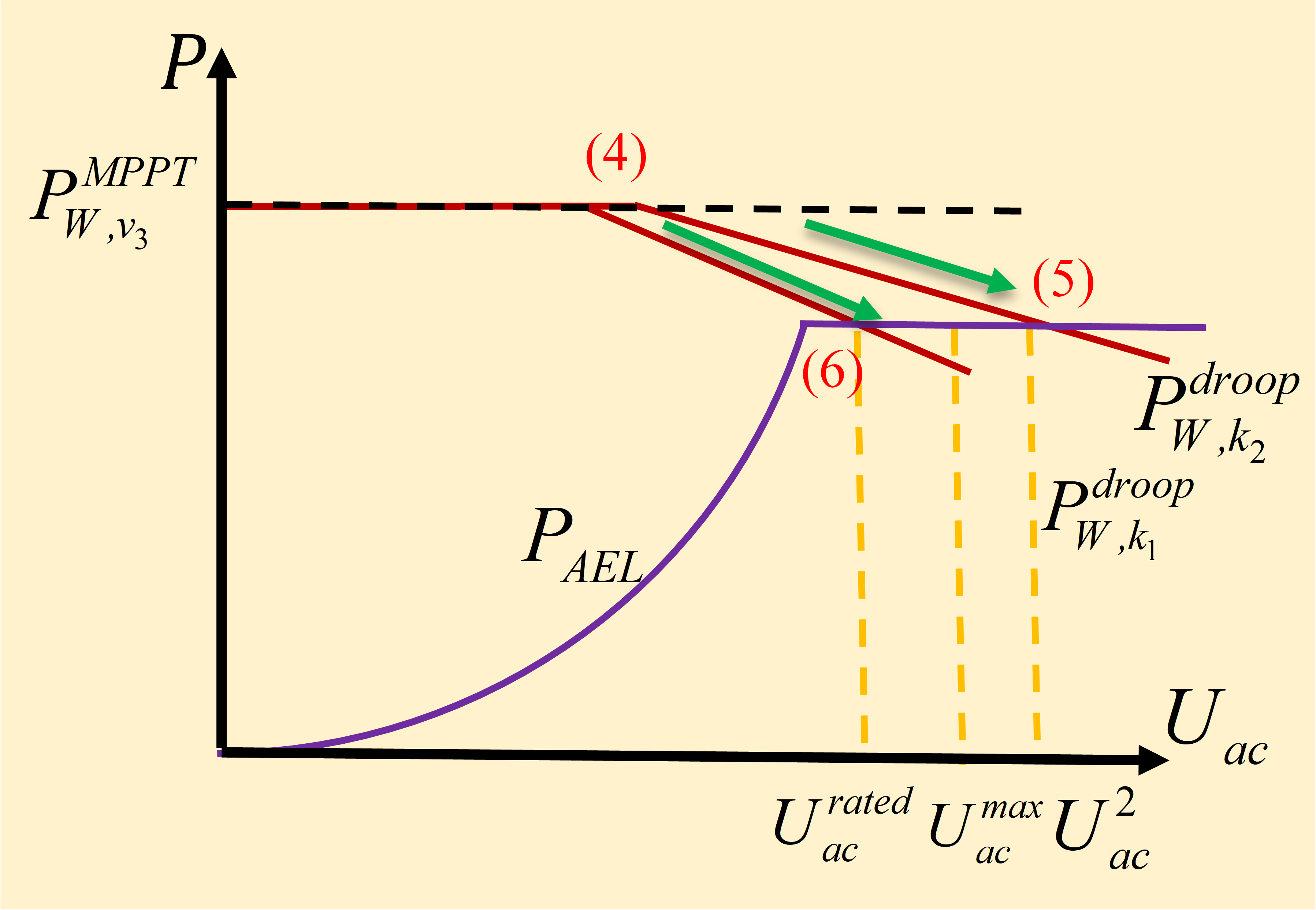}
       \end{minipage}
       \label{fig_5b}
   }
   \caption{The basis of voltage stability:(a) N-mode; (b) E-mode}
   \label{fig_5}
 \end{figure}
   The SWSE microgrid has two operation modes: normal (N-mode) and emergency (E-mode). In the following section, we introduce the characteristics and principles of voltage stability for these two operational modes.
 \subsubsection{N-mode}\ \par
 \textit{\textbf{a) The AEL converter and the AEL are collectively modeled as a resistive load}}
 
 The voltage across the AEL $U_{AEL}$ is founded by integrating equation (\refeq{eq3}) with equation (\refeq{eq5}):
 \begin{equation}
   \label{eq6}
   U_{A E L}=1.225 d U_{a c}
 \end{equation}
 
 Subsequently, the AEL power $P_{AEL}$ is given by:
 \begin{equation}
   \label{eq7}
   P_{A E L}=\frac{\left(2.34 d u_{a c}\right)^2}{R_{A E L}}=\frac{u_{a c}^2}{R_{A E L} /(2.34 d)^2}=\frac{u_{a c}^2}{R_{e q}}
 \end{equation}
 where $R_{AEL}$ is the equivalent resistance of AEL, and $R_{eq}$ is the equivalent resistance seen from the AC voltage bus.\par
 Equation (\refeq{eq7}) shows that the AEL power is a quadratic function of AC voltage's root mean square (RMS) value, depicted as purple solid curves in Fig. \ref{fig_5a}. It should be noted that variations in the AEL operating state lead to changes in $R_{eq}$, which modifies the slope of the \textit{P-U} curve. Alternatively, adjusting the duty cycle of the IPBC can actively change the slope of the \textit{P-U} curve.
 
 \textit{\textbf{b) DFIG can be modeled as a current source}}
 
 The DFIG output power follows the MPPT trace, as detailed in the equation (\refeq{eq1}). This output power solely depends on the wind speed and rotational speed of the WT, so DFIG \textit{P-U} characteristics are represented by dashed black lines parallel to the x-axis in Fig. \ref{fig_5a}, indicating that the DFIG output power is independent of the AC voltage. Let $P_{W, v_1}^{M P P T}$ and $P_{W, v_2}^{M P P T}$ denote the DFIG output power under MPPT mode at wind speeds $ v_{1} $ and $ v_{2} $, respectively. The rate of change of the DFIG output power is not only constrained by its own ramp-up/down limits but also by the ramp-up/down limits of the AEL:
 \begin{equation}
   \label{eq8}
   \left|\frac{P_W(t)-P_W(t-\Delta t)}{\Delta t}\right| \leq \min \left\{\delta_{A E L}, \delta_{wind }\right\}
 \end{equation}
 where $\delta_{A E L} $ and $\delta_{wind} $ are the maximum ramp-up/down rates of the AEL and DFIG, respectively. The operating boundaries of the N-mode are:
 \begin{equation}
   \label{eq9}
   P_{AEL}^{min}<P_W^{MPPT}<P_{AEL}^{rated}
 \end{equation}
 
 \textit{\textbf{c) The mechanism of voltage stability}}
 
 Fig. \ref{fig_5a} illustrates that the \textit{P-U} curves of DFIG and AEL always intersect, where the intersection point represents the microgrid system's stable operating point. Let point (1) be the initial operation point, with wind speed $ v_{1} $ and rated RMS AC voltage $U_{ac}^{rated} $. When the wind speed increases from $ v_{1} $ to $ v_{2} $, the stable operation point shifts from point (1) to point (2), with the voltage changing from $U_{ac}^{rated} $ to $U_{ac}^{1} $. To keep the voltage at the rated value and minimize its fluctuations, the duty cycle of the IPBC is quickly adjusted to actively alter the slope of the AEL \textit{P-U} curve. This changes the stable operation point from point (2) to point (3), and restores the voltage from $U_{ac}^{1} $ to $U_{ac}^{rated} $.
 \subsubsection{E-mode}\ \par
 To maximize the AEL utilization, the capacity of the DFIG is typically configured to exceed that of the AEL[32]. Consequently, when wind speed is exceedingly high, the MPPT power may be greater than the AEL's rated power. In this case, the AEL \textit{P-U} curve is capped at the rated power, no longer having an intersection with the DFIG \textit{P-U} curve, as shown in Fig. \ref{fig_5b}. This means that the AEL loses the voltage regulation capability, and the microgrid system no longer has a stable operation point. The E-mode is specifically designed to address this issue in this strong-wind scenario. The operating boundaries of the E-mode are:
 \begin{equation}
   \label{eq10}
   \begin{aligned}
   & P_W^{MPPT}<P_{AEL}^{min} \text { or } P_W^{MPPT}>P_{AEL}^{rated} \text { or } \\
   & \left|\frac{P_W(t)-P_W(t-\Delta t)}{\Delta t}\right| \geq \min \left\{\delta_{A E L}, \delta_{wind }\right\}
   \end{aligned}
 \end{equation}
 
 \textit{\textbf{a) AEL \textit{P-U} characteristic}}
 
 As elaborated above, in the E-mode, the AEL power reaches its rated value, and the AEL \textit{P-U} curve is capped at the rated power: any further increase in the voltage does not correspond to the rise in the AEL power, as shown in Fig. \ref{fig_5b}.
 
 \textit{\textbf{b) DFIG \textit{P-U} characteristic}}
 
 A drooping DFIG \textit{P-U} characteristic curve must be constructed to intersect with the AEL \textit{P-U} characteristic curve. In Fig. \ref{fig_5b}, $P_{W, k_1}^{d r o o p}$ and $P_{W, k_2}^{d r o o p}$ are two droop \textit{P-U} curves with different slopes $k_1 $ and $k_2 $, the DFIG power decreases as the AC bus voltage increases. This corresponds to the fact that the AC bus voltage exceeding the rated value indicates that DFIG power is higher than the AEL power. 
 
 \textit{\textbf{c) The mechanism of voltage stability}}
 
 Fig. \ref{fig_5} illustrates that in the E-mode, the DFIG power automatically changes in response to changes in the AC bus voltage to maintain voltage stability. Initially, the system operates in the N-mode, where the AC bus voltage is the rated value $U_{a c}^{rated }$. As wind speed increases, the DFIG power reaches to $P_{W, v_3}^{MPPT}$. At this point, the AEL cannot fully absorb the DFIG power, causing the bus voltage to rise and posing an overvoltage risk. When the central control system detects this situation, it switches to the E-mode shown in Fig. \ref{fig_5b}. The DFIG power $P_{W, k_2}^{droop}$ will decrease along the slope $k_2 $ until the system stabilizes at the intersection point (5). However, this stable point corresponds to a bus voltage exceeding the maximum limit, posing a risk of damaging electrical equipment. Therefore, it is necessary to design the droop curve carefully to ensure that the stable AC bus voltage point is between the rated voltage and the maximum limit. An example of a ``good'' stable point is the point (6), where the AC bus voltage falls between the rated value and the maximum limit.
 \subsection{The steady-state optimal scheduling model}
 A steady-state optimal scheduling model developed by Kang Ma is introduced, reflecting the stability principle introduced above.
 
 The objective of the SWSE microgrid is to maximize hydrogen production, which is equivalent to maximizing the AEL power:
 \begin{equation}
   \label{eq11}
   P_{A E L}^{result} \equiv \max P_{A E L}
 \end{equation}
 
 The following equations give constraints:
 \begin{equation}
   \label{eq12}
   P_{A E L}=P_W
 \end{equation}
 \begin{equation}
   \label{eq13}
   \frac{d P_{A E L}}{d t}=\frac{d P_W}{d t}
   \end{equation}
 \begin{equation}
   \label{eq14}
   \begin{aligned}
   P_{A E L}(t) & \leq P_{A E L}(t-\Delta t)+\min \left\{\delta_{A E L}, \delta_{wind}\right\} \Delta t \\
   & =P_{A E L}^{R C}(t)
   \end{aligned}
 \end{equation}
 \begin{equation}
   \label{eq15}
   \begin{cases}P_{A E L}^{min } \leq P_{A E L} \leq P_{A E L}^{rated},&\text { if } P_{A E L}^{rated}<P_{W, v_{wind}}^{M P P T} \\ P_{A E L}^{min } \leq P_{A E L} \leq P_{W, v_{wind}}^{M P P T},&\text { if } P_{A E L}^{min} \leq P_{W, v_{wind}}^{M P P T} \leq P_{A E L}^{rated} \\ P_{A E L}=0,&\text { if } P_{A E L}^{min}>P_{W, v_{wind}}^{M P P T}\end{cases}
 \end{equation}
 
 The SWSE microgrid power balance constraint is given by (\refeq{eq12}). According to constraint (\refeq{eq13}), the power change rate of DFIG and AEL should also be the same. The ramp-up/down limit of the AEL power is provided by (\refeq{eq14}). The AEL power range is provided by (\refeq{eq15}). It is found that:
 \begin{equation}
   \label{eq16}
   P_{A E L}(t)= \begin{cases}\min \left\{P_{W, v_{wind}}^{M P P T}, P_{A E L}^{rated}, P_{A E L}^{R C}(t)\right\}, \\\text { if } P_{A E L}^{min } \leq P_{W, v_{wind}}^{M P P T} \\ 0, \\\text { if } P_{A E L}^{min }>P_{W, v_{wind}}^{M P P T}\end{cases}
 \end{equation}
 
 When the maximum power that the DFIG can output at wind speed $V_{wind}$ is less than the lower limit of the AEL power, the AEL should be disconnected from the microgrid, and the DFIG should operate under a no-load condition. In other cases, the AEL power equals the minimum value among $P_{W, v_{wind}}^{M P P T}$, $P_{AEL}^{rated}$ and $P_{AEL}^{RC}(t)$. 
 
 \section{Control Logic fot the Microgrid}
 \begin{figure*}[t]
  \centering
  \includegraphics[scale=0.9]{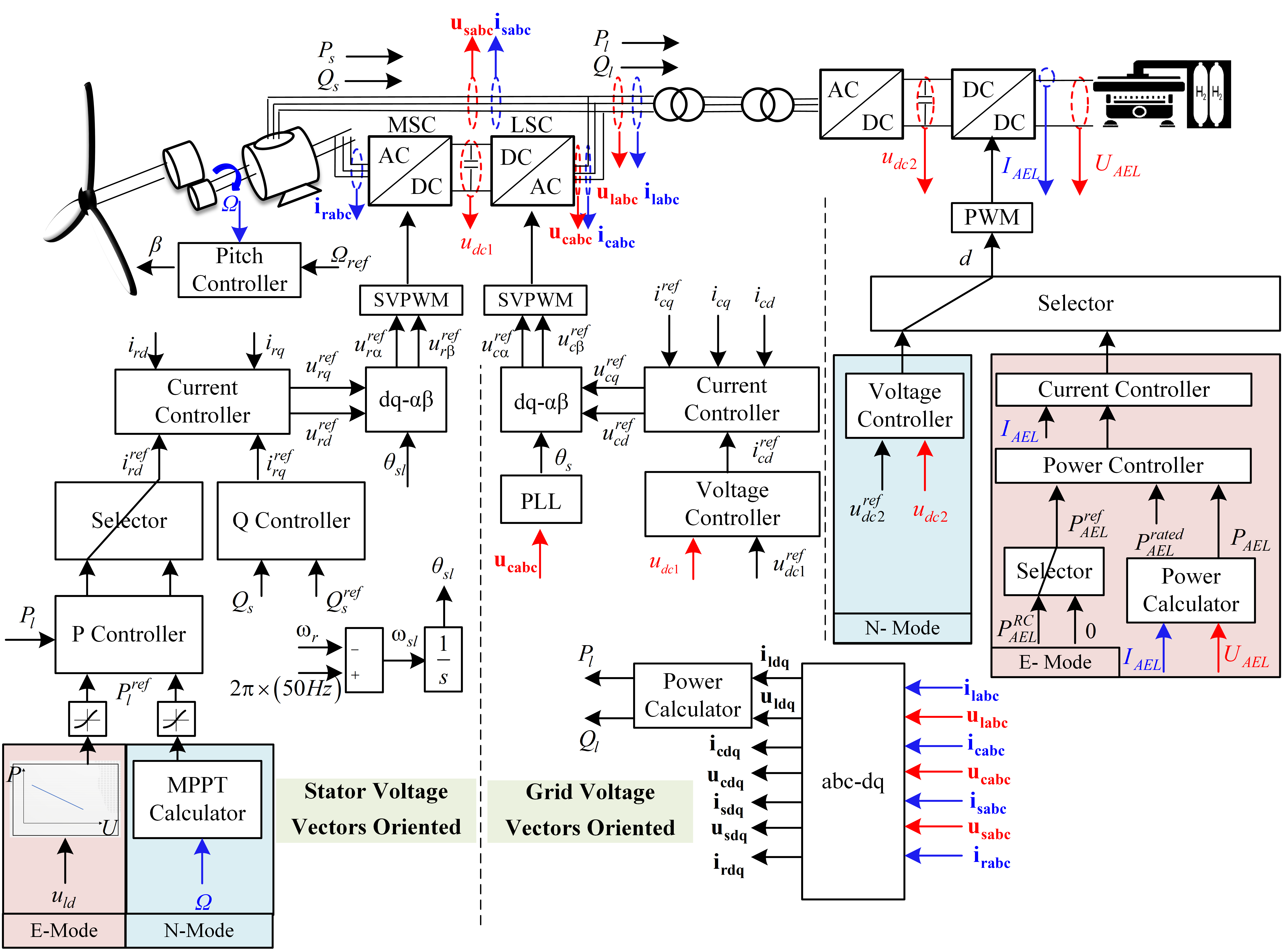}
  \caption{Overview of the SWSE microgrid model with control structure for the WT, MSC, LSC, and IPBC.}
  \label{fig_6}
\end{figure*}
 Fig. \ref{fig_6} depicts the overall control diagram based on the stability mechanisms and the solution to the SWSE microgrid steady-state operation model discussed in section III. The following sections will introduce the control blocks of each part.
 \subsection{Pitch Control}
 The pitch control restrains the WT rotational speed and power at maximum values when the wind speed is exceedingly high. The control block details are in Appendix \cite{ref24} as a well-established control scheme.
 \subsection{MSC Control}
 The purpose of the MSC control is to obtain the desired stator current, which determines the stator power. This study employs stator voltage orientation (SVO) \cite{ref37} as the control strategy for the MSC. The control block diagram comprises an outer power control loop and an inner current control loop using PI controllers. The control block details are in Appendix \cite{ref24}.
 
 The control block diagram is the same for both the N-mode and the E-mode, with the only difference being the reference value of the stator active power $P_{s}^{ref}$. Fig. \ref{fig_6} illustrates that under the N-mode, $P_{s}^{ref}$ is given by equation (\refeq{eq1}), and under the E-mode, $P_{s}^{ref}$ is given by:
 \begin{equation}
   \label{eq17}
   P_s^{ref}=P_{A E L}^{rated}+m\left(\sqrt{2} U_{a c}-\sqrt{2} U_{a c}^{rated}\right)
 \end{equation}
 where $\sqrt{2} U_{a c}$ equals to the line voltage in the d-axis when the SVO control is adopted and amplitude-preserving Park transformation is used; $m$ is a droop coefficient given by:
 \begin{equation}
   \label{eq18}
   m=\frac{P_{A E L}^{rated}-x P_{A E L}^{rated}}{\sqrt{2} U_{a c}^{rated}-\sqrt{2} U_{a c}^{max}}
 \end{equation}
 where $x $ determines the slope of the droop curve, and it ranges from 0 to 1.
 \subsection{LSC Control}
 The LSC control consists of an inner current control loop with fast dynamics and an outer DC voltage control loop with slower dynamics. The objectives of the LSC are to maintain the DC voltage at the reference value and to regulate the power factor. The control block details are in Appendix \cite{ref24}.
 
 It should be noted that the LSC control is the same for both the N-mode and the E-mode. 
 
 \subsection{IPBC Control}
 \begin{figure}[t]
   \centering
   \includegraphics[scale=1]{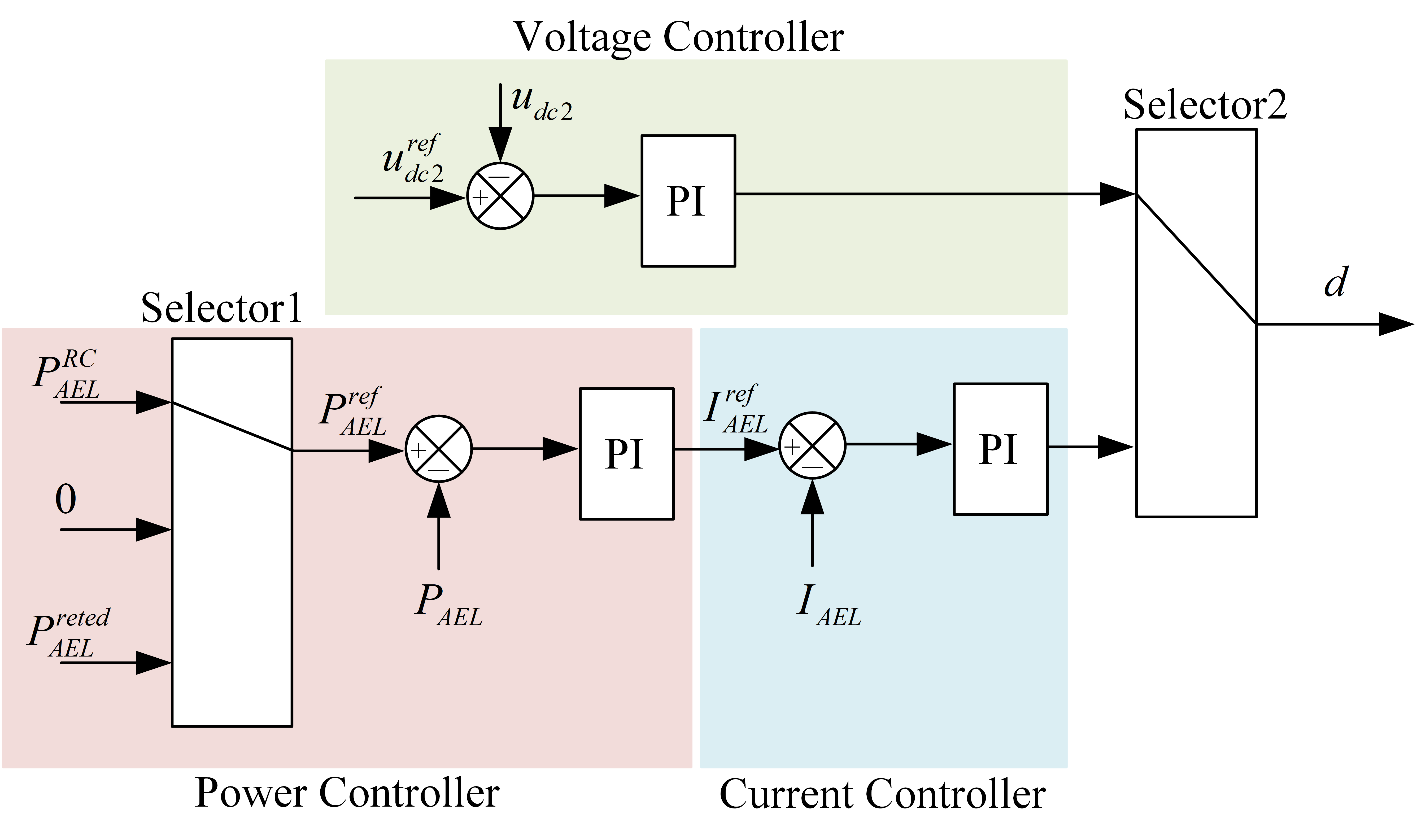}
   \caption{Control block diagram for the IPBC control}
   \label{fig_7}
 \end{figure}
 Fig. \ref{fig_7} illustrates the control block diagram of the IPBC. In the N-mode, the duty cycle $d$ is derived from the voltage controller based on the difference between the reference DC voltage $u_{d c 2}^{r e f}$ and the measured value $u_{d c 2}$ at the input side of the IPBC. When any of the conditions listed in equation (\refeq{eq10}) is satisfied, selector 2 switches from the voltage controller to the power-current controller, and the control block switches to the E-mode. When the ramp-up rate of AEL power is greater than the maximum ramp-up rate, the reference AEL power is set to $P_{A E L}^{R C}$ given by equation (\refeq{eq14}); when $P_{W,v_{wind}}^{MPPT}$ is lower than $P_{AEL}^{min}$, the AEL is disconnected with $P_{A E L}^{r e f}=0$; when $P_{W,v_{wind}}^{MPPT}$ exceeds $P_{A E L}^{rated}$, $P_{A E L}^{r e f}=P_{A E L}^{rated}$.
 
 \subsection{Control Mode Switching}
\begin{figure}[h]
   \centering
   \includegraphics[scale=1]{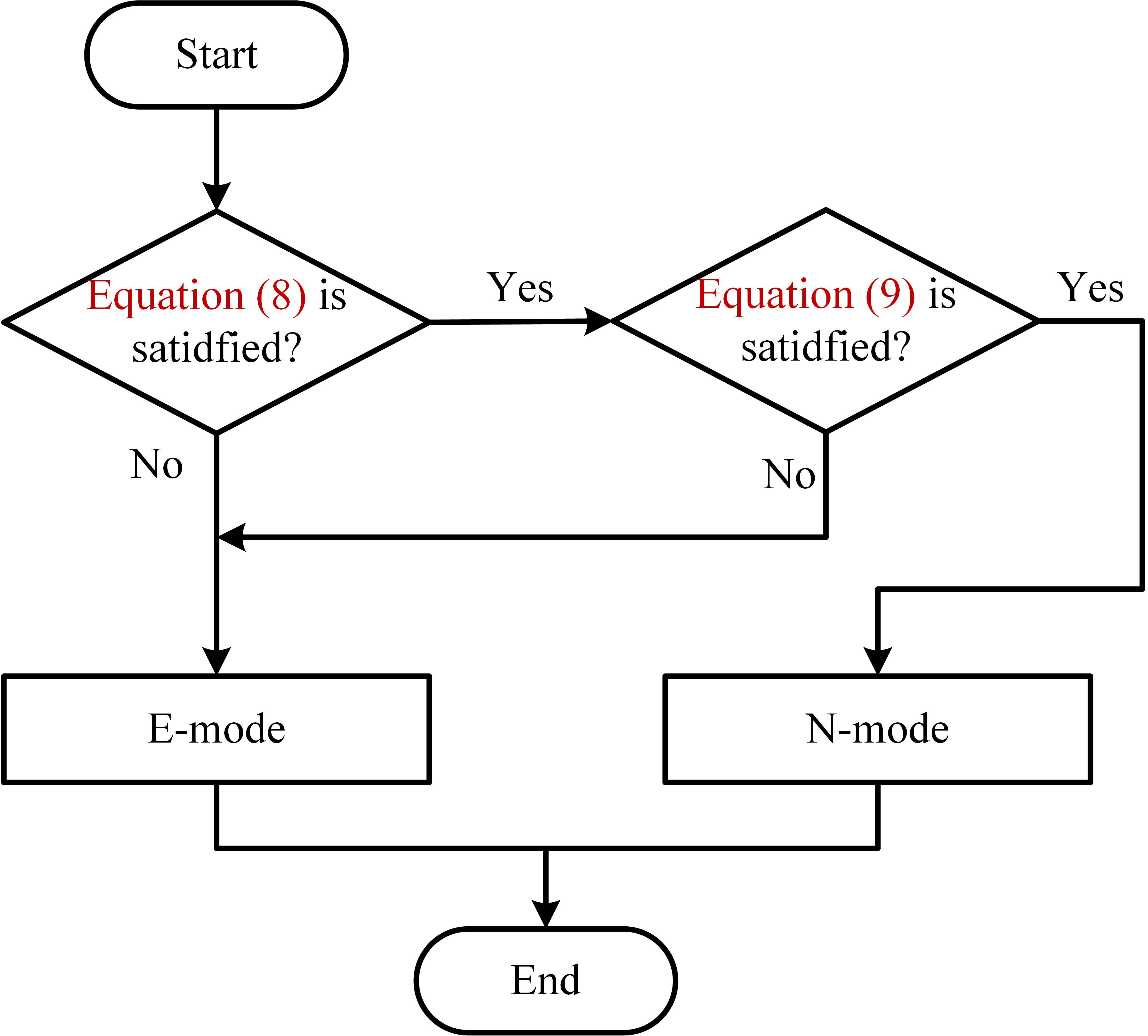}
   \caption{Control mode switching logic flowchart}
   \label{fig_8}
 \end{figure}

 The microgrid central control system receives status information from DFIG and AEL, continuously monitors the microgrid, and gives control mode switching commands if necessary. Fig. \ref{fig_8} illustrates that the microgrid operates in N-mode only when both equation (\refeq{eq8}) and equation (\refeq{eq9}) are satisfied. If either condition is violated, the microgrid transitions to E-mode to ensure stable and reliable operation.
 \section{Simulation Results}
 This section presents the SWSE microgrid's simulation results. The simulation was conducted in Matlab/Simulink R2022b; the simulation parameters are listed in Appendix \cite{ref24}. 
 
 Case studies are organized as follows: case study 1 validates N-Mode without considering the ramp-up/down rate constraint. Case study 2 validates the N-Mode while considering the ramp-up/down rate constraint. Case study 3 simulates a catastrophic scenario in the N-Mode, where the DFIG exceeds the AEL's rated power, but the control mode does not transition to the E-Mode. Case study 4 validates the transition between the E-Mode and the N-Mode. Case study 5 evaluates the performance of the SWSE microgrid control strategy under more challenging conditions using stochastic wind speed testing. 
 \subsection{Case Study 1}
 In this case, the AEL capacity is the same as the DFIG capacity; the ramp-up/down rate limit is not considered.
 \subsubsection{0-3 s} \ \par
 The first 0.5 s corresponds to the black-start process of the microgrid, which is beyond this paper's scope. The initial wind speed is 9 m/s, as shown in Fig. \ref{fig9} (a). The DFIG power is 0.48 p.u. and the AEL power is 0.4 p.u. with a total loss of approximately 17\%, as shown in Fig. \ref{fig9} (b).
 \subsubsection{3-7 s} \ \par
 The wind speed steps up to 11 m/s after 3 s, as shown in Fig.\ref{fig9} (a), causing the DFIG mechanical torque to increase from 0.5 p.u. to 0.8 p.u., but the electromagnetic torque increases gradually. The imbalance between the DFIG's torque increases the rotational speed from 0.5 p.u. to 0.75 p.u. Fig. 8 \ref{fig9} (b) shows that the DFIG power follows the MPPT reference power and increases up to 0.82 p.u. as the rotational speed increases. Under the control of the 16-IPBC, the power trajectory of the AEL aligns with that of the DFIG.
 \subsubsection{7-10 s}\ \par
 At 7s, the wind speed falls back to 9 m/s, causing the mechanical torque to decrease from 0.8 p.u. to 0.5 p.u., and the rotational speed hence decreases from 0.75 p.u. to 0.5 p.u. The DFIG power gradually decreases from 0.82 p.u. to 0.48 p.u., while the AEL power actively tracks the DFIG power.
 
 Fig. \ref{fig9} (b) illustrates the microgrid system frequency, which remains constant despite variations in DFIG power. This is fundamentally different from traditional power systems, where the frequency reflects the active power imbalance between generations and loads. Fig. \ref{fig9} (c) illustrates that the 16-IPBC and AEL maintain the microgrid's AC voltage at the rated value. 
 
 Case study 1 validates the control strategy for the SWSE microgrid without BESS in N-mode. It demonstrates that the converter on the AEL side can maintain power balance, ensuring microgrid voltage stability. Furthermore, the results confirm that the MSC can solely determine the microgrid frequency, unaffected by power fluctuations.    
 
 \subsection{Case Study 2}
 The AEL capacity is configured at a 1:1 ratio with the DFIG capacity. The wind speed is the same as in \mbox{case study 1}. Different from case study 1, the ramp-up/down constraint is considered.
 \subsubsection{0-3 s}\ \par
 The DFIG power ramp-up rate is limited at 5\% load/s, so neither the DFIG power nor the AEL power reaches a steady state within the first 3 s, and they continue to increase, as shown in Fig. \ref{fig9} (d).
 \subsubsection{3-7 s}\ \par
 The wind speed steps up to 11 m/s, and the imbalance between the DFIG's mechanical and electromagnetic torques causes the rotational speed to increase. However, since the ramp-up rate of the electromagnetic torque is limited, the torque balance can not be reached rapidly. Thus, the rotational speed $\mathit{\Omega}$ keeps increasing until it reaches its upper limit of 1.2 p.u. at 5.3 s, as shown in Fig. \ref{fig9} (d). Then, the pitch control described in section IV increases the pitch angle to reduce the mechanical torque, thereby limiting $\mathit{\Omega}$.
 \subsubsection{7-10 s}\ \par
\begin{figure}[t]
  \centering
  \includegraphics[scale=0.3]{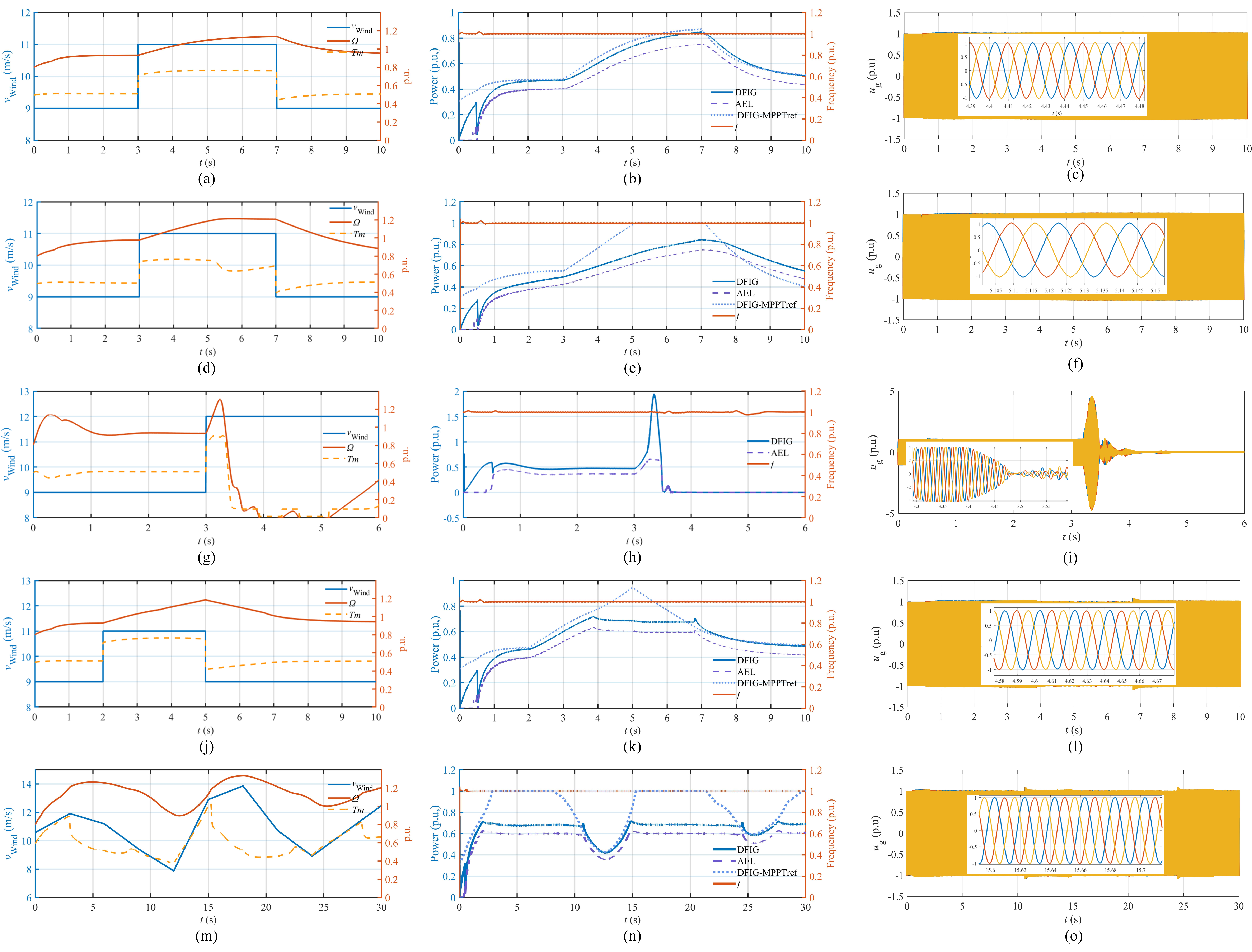}
  \caption{The simulation results of the SWSE microgrid}
  \label{fig9}
\end{figure}

 Fig. \ref{fig9} (d) shows the wind speed back to 9 m/s, causing the mechanical torque to decrease from 0.7 p.u. to 0.5 p.u. The mechanical torque falls below the electromagnetic torque, causing $\mathit{\Omega}$ to decrease. Fig. \ref{fig9} (e) shows that the DFIG power begins to decline. However, the ramp-down limit causes its response to lag behind the MPPT reference power.
 
 A comparison with the results of case study 1 reveals that the ramp-up/down constraint the rate of power change when the wind speed changes. When the wind speed increases rapidly, hydrogen production may decrease due to wind curtailment. Conversely, when the wind speed decreases rapidly, hydrogen production may increase. Fig. \ref{fig9} (e) and \ref{fig9} (f) illustrate that ramp-up/down constraint does not affect frequency and voltage stability.
 
 \subsection{Case Study 3}
 The AEL capacity is configured as 60\% of the DFIG capacity without considering the ramp-up/down constraint. Blocking the control mode switching commands issued by the microgrid central control system.
 \subsubsection{0-3 s}\ \par
 The microgrid operates in N-mode. During the first 3 s, the system operates in a steady state. 
 \subsubsection{3-6 s}\ \par
 The wind speed steps up from 9 m/s to 12 m/s at $t=3 \mathrm{~s}$. To balance the wind power, the AEL power rises until it reaches the rated power of 0.6 p.u. Fig. \ref{fig9} (h) shows that AEL power is capped at its rated value and cannot absorb more wind energy, leading to a voltage surge in the AC transmission line, as shown in  Fig. \ref{fig9} (i). Consequently, the protection system trips the DFIG and AEL, resulting in a system blackout. 
 
 Fig. \ref{fig9} (h) illustrates that as long as the MSC operates correctly, the frequency control method described in section III can effectively maintain the system frequency, regardless of the power balance between the DFIG and AEL.
 
  Case study 3 simulates a system failure scenario. The AEL power attains its rated value, and the DFIG reference power is calculated by the MPPT algorithm. The characteristic curves of the AEL and DFIG are parallel, with no intersection, meaning there is no stable operating point. This case highlights the necessity of the E-mode to prevent system failure.
 
 \subsection{Case Study 4}
 The AEL capacity is configured as 60\% of the DFIG capacity without considering the ramp-up/down rate constraint. Unlike case study 3, the control mode switching command is effective in this case.
 \subsubsection{0-2 s}\ \par
 The initial wind speed is 9 m/s, and the SWSE system operates in N-mode.
 \subsubsection{2-5 s}\ \par
 The wind speed steps up to 11 m/s at $t=2 \mathrm{~s}$, and AEL power reaches the rated power of 0.6 p.u. at $t=3.8 \mathrm{~s}$. Then, the microgrid central control system sends control mode switching commands to the DFIG and AEL controllers, and the SWSE microgrid operation mode switches from the N-mode to the E-mode. The power reference for the MSC power control loop is now determined by the  droop curve given by equation (\refeq{eq17}), instead of the MPPT algorithm, as shown in Fig. \ref{fig9} (k). A new power balance between the DFIG and the AEL is achieved through wind power curtailment, thus maintaining voltage stability, as shown in Fig. \ref{fig9} (l).
 \subsubsection{5-10 s}\ \par
 The wind speed falls from 11 m/s to 9 m/s at $t=5 \mathrm{~s}$, the mechanical torque is less than the electromagnetic torque, leading to decreased rotational speed as shown in Fig. \ref{fig9} (j). However, because the DFIG MPPT reference power remains higher than the rated AEL power, the system continues to operate in the E-mode. The rotational speed decreases to the point where $P_{W, v_{w i n d}}^{M P P T}$ becomes less than $P_{AEL}^{rated}$ at 6.7 s, when the WT can no longer capture sufficient wind energy to sustain the DFIG power. To prevent the DFIG from halting, the operating mode switches back to the N-mode.
 Fig. \ref{fig9} (l) shows that in the E-mode, the system can operate smoothly through emergency operating conditions when the wind speed is high.
 
 \subsection{Case Study 5}
 This case employed the Weibull distribution model and stochastic wind speed generation method to simulate the operation of the SWSE microgrid under more complex and realistic conditions, which correspond to high wind conditions. As shown in Fig. \ref{fig9} (m), the average wind speed is 10.5 m/s, fluctuating in a wave-like pattern.
 
 Fig. \ref{fig9} (m) illustrates that during the 30-second simulation period, the microgrid underwent five control mode switches. The electrolyzer operates at its rated power during high wind speeds, while MPPT functionality is implemented during low wind speeds. For instance, at $t=2.4 \sim 10.2 \mathrm{~s}$, DFIG operates under droop control. Throughout this process, both frequency and voltage remain stable. 
 
 This case demonstrates that the proposed SWSE microgrid and its coordinated control strategy can maximize hydrogen production while ensuring system safety, without using BESS for grid forming.
 \section{Conclusion}
 
 As the scale of renewable energy installations continues to expand, accommodating surplus energy is becoming increasingly critical. Green hydrogen production from renewable energy offers a promising solution to this issue. However, the reliance on grid-forming BESS raises the LCOH. This study proposes a novel control strategy for a DFIG supplying an AEL in an islanded storage-less microgrid to address this challenge.
 
 The SWSE system comprises a WT, DFIG, B2B converter, 24-UR, 16-IPBC, and AEL. This study introduces an innovative 24-UR model where the AC side is a CS rather than a VS, different from the conventional model widely presented in textbooks and scholarly works on power electronics.
 
 In the conventional power system principle, grid frequency is closely linked to active power balance. This study challenges this traditional principle by fixing the microgrid frequency at 50 Hz through modulation wave control of the MSC. This 100\% converter-based storage-less microgrid treats frequency as a virtual parameter entirely decoupled from the active power balance. This provides a new perspective on frequency control in islanded microgrids with only converter-based sources. Instead of frequency, the AC voltage magnitude reflects the active power balance in the microgrid. The \textit{P-U} characteristic curve illustrates the relationship between the active power and the AC voltage, uncovering the fundamental mechanism ensuring a stable operating voltage point for the microgrid. In the N-mode, DFIG power tracks the MTTP trace. At the same time, the voltage is stabilized by dynamically adjusting the equivalent resistance on the hydrogen production side via IPBC, thereby modifying the AEL \textit{P-U} characteristic curve. In the E-mode, the DFIG power tracks the \textit{P-U} droop curve and curtails the wind power to maintain a constant AC voltage.
 
 The primary contribution of this research lies in developing a novel control strategy for a DFIG supplying an AEL in an islanded storage-less microgrid. The WT controller constrains the rotational speed and mechanical power of the WT to their maximum values under overrated wind speed conditions. The LSC controller is designed to maintain the DC-link voltage at a reference value while regulating the power factor. The MSC controller governs the DFIG active power, following the MPPT trajectory in the N-mode and the droop curve in the E-mode. In the N-mode, the IPBC stabilizes the microgrid's AC voltage amplitude by maintaining the input DC voltage of the IPBC at a fixed value. In the E-mode, the controller switches to the power control loop to satisfy the AEL power operating range and ramp-up/down constraints.
 
 This study validates the proposed control strategy through five distinct simulation scenarios, demonstrating its effectiveness in enabling the SWSE microgrid to operate reliably under various conditions. The results confirm that the strategy not only maximizes hydrogen production but also ensures the microgrid's safe and stable operation.
\bibliographystyle{IEEEtran}  
\bibliography{refs}

\end{document}